\documentclass[a4paper]{jpconf}
\usepackage{graphicx,amsmath}
\usepackage{hyperref}
\usepackage{youngtab}

\newcommand{\be}{\begin{equation}}
\newcommand{\ee}{\end{equation}}
\newcommand{\bea}{\begin{eqnarray}}
\newcommand{\eea}{\end{eqnarray}}

\begin{document}
\title{A worldline approach to colored particles}

\author{Fiorenzo Bastianelli$^1$, Roberto Bonezzi$^1$, Olindo Corradini$^{2,3}$, Emanuele Latini$^4$ and
Khaled Hassan Ould-Lahoucine$^5$}

\address{$^1$ Dipartimento di Fisica ed Astronomia, Universit{\`a} di Bologna and\\
INFN, Sezione di Bologna, via Irnerio 46, I-40126 Bologna, Italy\\
$^2$ Facultad de Ciencias en F\'isica y Matem\'aticas,
Universidad Aut\'onoma de Chiapas, Ciudad Universitaria, Tuxtla Guti\'errez 29050, M\'exico\\
$^3$ Dipartimento di Scienze Fisiche, Informatiche e Matematiche,\\ Universit\`a di Modena e Reggio Emilia, Via Campi 213/A, I-41125 Modena, Italy\\
$^4$ Institut f{\"u}r Mathematik, Universit{\"a}t Z{\"u}rich-Irchel, Winterthurerstrasse 190, CH-8057 Z{\"u}rich, Switzerland\\
$^5$ D\'epartement de Physique, Universit\'e Setif 1, Algeria}

\begin{abstract}
Relativistic particle actions are a useful tool to describe quantum field theory effective actions using a string-inspired first-quantized approach. Here we describe how to employ suitable particle actions in the computation of the scalar contribution to the one-loop gluon effective action. 
We use the well-known method of introducing auxiliary variables that create the color degrees of freedom. 
In a path integral they implement automatically the path ordering needed to ensure gauge invariance. 
It is known that the color degrees of freedom introduced this way form a reducible
representation of the gauge group. We describe a method of projecting onto the fundamental representation (or any other chosen 
irrep, if desired) of the gauge group. Previously, we have discussed the case of anticommuting  auxiliary variables. 
Choosing them to be in the fundamental representation allows to obtain, without any extra effort, also the situation in which  
the color is given by any {\it antisymmetric} tensor product  of the fundamental. 
Here, we describe the novel case of bosonic auxiliary variables. They can  be used equivalently for creating the color charges in 
 the fundamental representation.  In addition one gets, as a byproduct, the cases where the particle can have the color  
 sitting in any {\it symmetric} tensor product of the fundamental.
This is obtained by tuning to a different value a Chern Simons coupling, present in the model, which controls how  the projection is achieved.
 \end{abstract}

\section{Introduction}
The worldline formalism has proved to be an efficient alternative to conventional second-quantized methods to compute Quantum Field Theory amplitudes in the presence of external fields, such as (non)abelian vector fields~\cite{Schubert:2001he} and/or gravity~\cite{Bastianelli:2002fv, Bastianelli:2002qw},
by using suitable quantum-mechanical path integrals.
Within such formalism, spinning particle actions are helpful tools to describe the propagation of particles with spin. These models are obtained by implementing local supersymmetries on the worldline: the susy partners of the particle coordinates---upon quantization---introduce the spinorial degrees of freedom in the particle's Hilbert space, whereas the gauge fields of local supersymmetries impose Gupta-Bleuler type of constraints on the wave function, that turn out to be equation of motion of the second-quantized field operators~\cite{Brink:1976sz}. For free theories this method can be applied to particles of arbitrary spin~\cite{Gershun:1979fb}.  

Particles with spin can as well be described in terms of purely bosonic worldline models that involve suitable matrix-valued potential. However, already for the simplest abelian case, they require to path order the exponential of the classical action sitting inside
the path integral in order to guarantee gauge invariance and Lorentz invariance~\cite{Feynman:1951gn}. Spinning particle models, avoiding such complication, render the computation of scattering amplitudes much simpler.

 A similar path can be undertaken when a coupling to nonabelian gauge fields is considered. In such a case  color degrees of freedom need to be added to the particle's Hilbert space. One possibility is of course to couple the particle coordinates to an algebra-valued potential, and path order the associated path integral to guarantee nonabelian gauge invariance. This method was used for example in \cite{Bastianelli:1992ct} to compute trace anomalies using quantum mechanical models.
On the other hand one may as well reproduce the color degrees of freedom by the quantization of suitable (commuting or anticommuting) auxiliary variables
\cite{Balachandran:1976ya, Barducci:1976xq}. When used in a path integral they automatically implement the path ordering. 
However the quantization of such auxiliary variables yields the degrees of freedom of a reducible representation of the nonabelian gauge group.
Choosing the variables to be of the Grassmann type, i.e. anticommuting, yields a finite dimensional representation of the color group.
For this reason they were often preferred over the commuting ones. Nevertheless, the representation they create is reducible. 
For example, choosing the variables to be in the fundamental representation (antifundamental for the complex conjugates) 
one finds a space containing beyond the fundamental  all of its possible antisymmetric tensor products.
Choosing instead bosonic variables to start with, again in the fundamental representation, one finds an infinite-dimensional representation 
that splits into all possible symmetric tensor products of the fundamental. In either case the total representation is reducible.
Of course, one could use auxiliary variables in any other arbitrarily chosen
representation, and find in the corresponding Hilbert space all possible antisymmetric
or symmetric tensor products of the chosen representation.

One way of projecting onto a chosen irreducible representation is to couple the above auxiliary variables to a $U(1)$ worldline gauge field, with an additional Chern-Simons term, whose coupling constant is tuned to pick-out only the desired representation \cite{Bastianelli:2013pta}.
Indeed, the $U(1)$ gauge field has no dynamics in one dimension, and it acts just as a Lagrange multiplier that imposes
a constraint: it fixes the  occupation number of the auxiliary variables to a given value.
The latter is quantized quantum mechanically, and can be tuned by an appropriate Chern-Simons
coupling.  Selecting the Chern-Simons coupling to pick up only the states with occupation number one, i.e. the single particle states in the Fock space of the auxiliary variables, one keeps only the representation of the gauge group chosen for those auxiliary variables.
Modifying the value of the  Chern-Simons coupling, one picks up the states with different occupation number,
sitting in the antisymmetric (for anticommuting variables) or symmetric (for commuting variables) tensor products
of the representation chosen for the single particle states. For definiteness,  we will choose here
the fundamental of the $SU(N)$ gauge group.

 In the following we review such a construction and apply it to study the one-loop effective action of bosonic (spin 0)
and fermionic (spin $\tfrac12$) fields coupled to an external nonabelian field: this is related to the computation of the corresponding worldline path integral on the circle. Ideally, one would like to treat similarly the gluon itself in a first quantized way.
Dealing with its self-coupling is not straightforward, if one wishes to use a spinning particle model with ${\mathcal N}=2$ supersymmetry. However 
a method of introducing the coupling in that model has been introduced in \cite{Dai:2008bh}, and one might 
apply the projection described here for the color charges of that model as well.


\section{Scalar particle with color degrees of freedom}
The description of a free relativistic scalar particle can be obtained by simply implementing the constraint $p^2+m^2=0$. Classically this simply amounts to the relativistic dispersion relation between energy and momentum. Quantum-mechanically it must be interpreted as an operatorial equation, i.e. as a condition imposed on the particle quantum state, $(\hat p^2+m^2)|\phi\rangle=0$. From a path-integral view-point one describes the propagation of the relativistic scalar particle in terms of 
\begin{equation}
\int \frac{Dx Dp De}{\rm Vol~(Gauge)} e^{iS[x,p,e]}
\end{equation}
with the phase-space action given by
\begin{equation}
S[x,p,e] = \int_0^1 d\tau \Big[ p\cdot \dot x -\frac{e}{2}(p^2+m^2)\Big]~. 
\end{equation}  
The latter action is invariant under worldline diffeomorphisms and the einbein $e$ is the gauge field for such local symmetry, and it is necessary to divide out gauge equivalent configurations in the path integral. The algebraic equation of motion of the einbein implements the relativistic constraint. 

In order to minimally couple the scalar particle to a (non)abelian gauge field one replaces the conjugate momentum with the covariant momentum $\pi_\mu = p_\mu -g A_\mu(x)$ where $A_\mu(x) = A_\mu^a(x) T^a$ is an algebra-valued potential. The generators
$T^a$ are chosen in the representation that gives the non abelian charge of the scalar particle. For definiteness we choose
 $SU(N)$ as gauge group, and consider the fundamental representation.

 By integrating out the momenta, and performing a Wick rotation, one obtains the configuration space action
\begin{equation}
S[x,e;A] = \int_0^1 d\tau \Biggl( \frac{\dot x^2}{2e} -ig \dot x \cdot A +\frac{e m^2}{2}\Biggr) 
\end{equation}
and the associated  path-integral reads
\begin{equation}
\int \frac{Dx De}{\rm Vol~(Gauge)} ~{\cal P} e^{-S[x,e;A]}
\end{equation}
where a path-ordering prescription ${\cal P}$ has been inserted to guarantee gauge-invariance. Such ordering can be implemented by adding to the particle action some complex (commuting or anticommuting) auxiliary fields that sit in the (anti)fundamental representation of the gauge group. Hence  
\begin{equation}
S[x,c,\bar c, e; A] = \int_0^1 d\tau \Biggl[ \frac{\dot x^2}{2e}  +\frac{e m^2}{2} +\bar c^\alpha \dot c_\alpha-ig \dot x^\mu A^a_\mu\, \bar c^\alpha (T^a)_\alpha{}^\beta c_\beta\Biggr]\,,\quad \alpha=1,\dots,N~. 
\end{equation}
Upon canonical quantization,  the auxiliary variables $\bar c^\alpha$ and 
$c_\alpha$ become the creation and annihilation operators  $\hat c^{\dagger\alpha}$ and $\hat c_\alpha$. 
The related Fock space contains tensor products of fundamental representation.
One may represent such space in a coherent state basis $\langle \bar c^\alpha|$, left eigenstates of the creation operators $\hat c^{\dagger\alpha}$. On wave functions, $\hat c^{\dagger\alpha}$ act multiplicatively, whereas $\hat c_\alpha$ act as derivatives. Indeed, for a generic state $|\phi\rangle$  with wave function $\phi(x,\bar c^\alpha) = \langle \bar c^\alpha, x| \phi \rangle$ one has
\begin{eqnarray*}
&& \langle \bar c^\alpha, x| \hat c^{\dagger \alpha} |\phi \rangle = \bar c^\alpha \phi(x,\bar c^\alpha) \\
&& \langle \bar c^\alpha, x| \hat c_\alpha |\phi \rangle = \frac{\partial}{\partial \bar c^\alpha} \phi(x,\bar c^\alpha)
\end{eqnarray*}
and the wave function $\phi(x,\bar c^\alpha)$ can be written as a $\bar c$-graded sum of fields
\begin{equation}\label{eq:expansion}
\phi(x,\bar c^\alpha) = \phi(x) + \bar c^\alpha \phi_\alpha(x) +\frac{1}{2!} \bar c^{\alpha_1} \bar c^{\alpha_2}\phi_{\alpha_1 \alpha_2} (x) +\cdots~,
\end{equation}
where the term $\phi_{\alpha_1\cdots\alpha_r}(x)$ is a field that sits in an symmetrized or  antisymmetrized 
tensor product of $r$ fundamental representations: the tensor is symmetric if the auxiliary variables are commuting, and it antisymmetric in the opposite case.  The generic scalar field described by~(\ref{eq:expansion}) thus is not, a priori, described by an irreducible representation of the gauge group. In fact the above sum is a finite sum if the auxiliary fields are anticommuting, whereas it is even an infinite sum if they are commuting. However, in both cases, one may impose a constraint 
\begin{equation}
\Big(\bar c^\alpha\frac{\partial}{\partial \bar c^\alpha} -r\Big) \phi(x,\bar c^\alpha)=0
\end{equation}    
in order to select only the field $\phi_{\alpha_1\cdots \alpha_r} (x)$ that has $n$ indices in the fundamental representation.  For example, for  a scalar field in the fundamental representation one fixes $r=1$. The latter constraint can be implemented in the worldline path integral by coupling the auxiliary variables to a $U(1)$ gauge field $a$, with the addition of Chern-Simons term whose coupling is $s=r\mp N/2$, i.e.
\begin{equation}
S[c,\bar c,a] = \int_0^1 d\tau \Big( \bar c^\alpha (\partial_\tau+ia)c_\alpha -isa\Big)~.
\end{equation} 
The factor $N/2$ solves an ordering ambiguity: the $a$ equation of motion $\bar c^\alpha c_\alpha -s=0$ corresponds,
by choosing a graded symmetric ordering,
to the operatorial equation $\frac12 (\bar c^\alpha \frac{\partial}{\partial \bar c^\alpha}\mp   \frac{\partial}{\partial \bar c^\alpha} \bar c^\alpha )-s=\bar c^\alpha \frac{\partial}{\partial \bar c^\alpha} -r =0$. Here the (upper) lower sign refers to (anti)commuting auxiliary fields.  
The way we are going to compute the path integral corresponds precisely to the ordering chosen above.
 This method was already employed in other worldline applications, such as for instance one-loop effective actions of abelian differential forms~\cite{Howe:1989vn,Bastianelli:2005vk,Bastianelli:2011pe}, higher-spin fields~\cite{Bastianelli:2009eh}, and quantum $(p,q)$-forms~\cite{Bastianelli:2012nh}. In~\cite{Bastianelli:2013pta} we treated the case of colored scalar and spin $\tfrac12$ particles with anticommuting auxiliary fields, and thus we dedicate the rest of the present manuscript to the case of commuting fields.

\subsection{The scalar QCD effective action and one-loop gluon amplitudes}
Equipped with the premises above, the one-loop QCD effective action generated by a colored scalar field coupled to a non-abelian gauge field, can be represented in terms of a colored scalar particle path integral defined on a circle  
\begin{equation}
\Gamma[A] = \int_{S^1} \frac{Dx D\bar c D c De Da}{\rm Vol~(Gauge)}~e^{-S[x,c,\bar c,e,a;A]}
\end{equation}
where the action is given by
\begin{equation}
S[x,c,\bar c,e,a;A] = \int_0^1d\tau \Biggl[ \frac{\dot x^2}{2e}  +\frac{e m^2}{2} -ig \dot x^\mu A^a_\mu\, \bar c^\alpha (T^a)_\alpha{}^\beta c_\beta+\bar c^\alpha (\partial_\tau+ia)c_\alpha -isa\Biggr]
\label{eq:W-action}
\end{equation}
that is invariant under local one-dimensional diffeomorphisms with gauge field $e(\tau)$, and local $U(1)$ symmetry with gauge field $a(\tau)$. One may thus gauge fix $e(\tau)=2T$, with $T$ being the conventional Fock-Schwinger time, and $a(\tau)=\theta$, with $\theta$ being an angle that parametrizes nonequivalent Wilson lines $z=e^{i\int_0^1 d\tau a} = e^{i\theta}$. Finally, the particle coordinates are taken with periodic boundary conditions. As mentioned above, we consider commuting auxiliary variables---i.e. fields that parametrize symmetric products of fundamental representations.  They are integrated on the circle with periodic boundary conditions, $c(1)=c(0)$ and $\bar c(1)=\bar c(0)$.  Hence,
\begin{equation}
\Gamma[A] =-\int_0^\infty\frac{dT}{T}e^{-Tm^2}\int_0^{2\pi}\frac{d\theta}{2\pi} e^{is\theta}\int_{PBC}Dx\int_{PBC}D\bar c D c~e^{-S[x,c,\bar c, 2T, \theta;A]}
\end{equation}  
and
\begin{equation}
S[x,c,\bar c,2T,\theta;A] = \int_0^1d\tau \Biggl[ \frac{\dot x^2}{4T}   -ig \dot x^\mu A^a_\mu\, \bar c^\alpha (T^a)_\alpha{}^\beta c_\beta+\bar c^\alpha (\partial_\tau+i\theta)c_\alpha\Biggr]
\end{equation}  
where the constant part $-Tm^2+is\theta$ has been factored out. Notice that the coupling with the $U(1)$ gauge field can be absorbed in the auxiliary fields by a field redefinition that implies a twist of the boundary conditions: $c(1)=e^{i\theta}c(0)$ and $\bar c(1)= e^{-i\theta}\bar c(0)$.  We will refer to the latter as ``twisted boundary conditions" (TBC). Therefore the above path integral involves an integral over nonequivalent TBC's that works as a projector on an  irreducible representation of the gauge group: it generalizes the approach of~\cite{D'Hoker:1995bj} where a sum over angles was instead used. 

The expression above is the starting point to compute a closed formula for the one-loop scalar contribution to gluon amplitudes. For such purpose we write the background nonabelian field as a sum of $n$ external gluons, with momenta $k_l$, polarizations $\varepsilon(k_l)$, and gluon colors $a_l$
\begin{equation}
A_\mu(x) = \sum_{l=1}^n \varepsilon_\mu(k_l)T^{a_l} e^{ik_l\cdot x}
\end{equation}
choosing the generators $T^a$ in the fundamental representation. One may thus extract the corresponding amplitude from the above path integral, by looking at terms linear in all the polarizations. This amounts to taking the path integral average of vertex operators
\begin{eqnarray}
&&\Gamma_{\rm scal}(k_1,\varepsilon_1,a_1;\dots;k_n,\varepsilon_n,a_n) =-(ig)^n \int_0^\infty\frac{dT}{T}e^{-Tm^2}\int_0^{2\pi}\frac{d\theta}{2\pi} e^{is\theta}\nonumber\\
&&\hskip4cm\times\int_{PBC}Dx\int_{TBC}D\bar c D c~e^{-S_2[x,c,\bar c]}\prod_{l=1}^n V_{\rm scal}[k_l,\varepsilon_l,a_l]\label{eq:G-amp}
\end{eqnarray}
where the ``gluon vertex operator" can be read off from the above action, and is given by
\begin{equation}
V_{\rm scal}[k_l,\varepsilon_l,a_l] =\int_0^1 d\tau~\bar c(\tau) T^{a_l} c(\tau)~e^{ik_l\cdot x(\tau) +\varepsilon_l\cdot \dot x(\tau)}\Bigr|_{\rm lin~\varepsilon}
\end{equation}
where only the linear term in $\varepsilon_l$ must be retained. Above the kinetic action reads
\begin{equation}
S_2[x,c,\bar c] = \int_0^1d\tau \Big( \frac{1}{4T} \dot x^2 +\bar c^\alpha \dot c_\alpha\Big)
\end{equation}
and it involves a zero-mode in the periodic $x$ sector. In order to invert the kinetic operator we need to factor it out. One way to do it is to remove the average of the closed path $x_0 =\int_0^1 d\tau x(\tau)$ and split $x(\tau) = x_0 +y(\tau)$, with $y(\tau)$ being periodic with vanishing average. We thus have $Dx = d^Dx_0~Dy$, and the free path integral normalization reads (we give some more details in the~\ref{appA})
\begin{equation}
\int_{PBC}Dy\int_{TBC}D\bar c D c~e^{-S_2[y,c,\bar c]} = (4\pi T)^{-D/2} ({\rm Det}_{TBC} (\partial_\tau))^{-N}= (4\pi T)^{-D/2} \left(2i\sin\frac{\theta}{2} \right)^{-N}
\end{equation}
whereas the propagators are
\begin{eqnarray}
 \langle y^\mu(\tau) y^\nu(\sigma)\rangle =-T\delta^{\mu\nu} G(\tau-\sigma)\,,\quad \langle c_\alpha(\tau) \bar c^\beta(\sigma) \rangle = \delta_\alpha^\beta \Delta(\tau-\sigma;\theta)
\end{eqnarray}
where
\begin{eqnarray}
&&G(\tau-\sigma) = | \tau-\sigma |-(\tau-\sigma)^2\\
&&\Delta(\tau-\sigma,\theta) =\frac{1}{2i\sin\frac{\theta}{2}}\left[ e^{i\frac{\theta}{2}}\theta(\tau-\sigma)+e^{-i\frac{\theta}{2}}\theta(\sigma-\tau)\right]~.
\end{eqnarray}
The amplitude~(\ref{eq:G-amp}) thus reduces to
\begin{eqnarray}
&&\Gamma_{\rm scal}(k_1,\varepsilon_1,a_1;\dots;k_n,\varepsilon_n,a_n)=\nonumber\\&& 
-(ig)^n (2\pi)^D\delta(\sum k_l) \int_0^\infty\frac{dT}{T}\frac{e^{-Tm^2}}{(4\pi T)^{D/2}}\int_0^{2\pi}\frac{d\theta}{2\pi} \frac{e^{is\theta}}{(2i\sin\frac{\theta}{2})^{N}} \prod_{l=1}^n\int_0^1 d\tau_l\nonumber\\
&& \times \Big\langle \prod_{l=1}^n {\cal V}_{c}(a_l;\tau_l)\Big \rangle ~\Big\langle \prod_{l=1}^n {\cal V}_{y}(k_l,\varepsilon_l;\tau_l)\Big \rangle \Big|_{{\rm m.l.}\, \varepsilon }
\end{eqnarray}
where
\begin{eqnarray}
&&{\cal V}_{c}(a;\tau) = (T^{a})_{\alpha}{}^{\beta} \bar c^{\alpha} (\tau) c_{\beta}(\tau)\\
&& {\cal V}_{y}(k,\varepsilon;\tau) = e^{ik\cdot y(\tau) +\varepsilon\cdot \dot y(\tau)}\label{eq:V-scal}
\end{eqnarray}
represent the (unintegrated) parts of the vertex operators involving auxiliary fields and coordinate fields respectively. The latter (cfr.~eq.(\ref{eq:V-scal})) is just the unintegrated vertex operator for scalar QED, and thus, from the last line of equation~(\ref{eq:G-amp}), one notices that---at the unintegrated level---the color part factores out. Also, the subscript ``m.l. $\varepsilon$" it to remind that one has to retain only the multilinear terms in the all $\varepsilon$'s. The momentum conservation Dirac delta-function in the amplitude above is generated by the integral over the zero mode $x_0$. The vacuum expectation value of~(\ref{eq:V-scal}) is the one that appears in the well known Bern-Kosower formula~\cite{Bern:1991aq} for scalar QED, i.e.
\begin{equation}
\Big\langle \prod_{l=1}^n {\cal V}_{y}(k_l,\varepsilon_l;\tau_l)\Big \rangle \Big|_{{\rm m.l.}\, \varepsilon } = \exp\Biggl\{T\sum_{ll'}\Biggl(k_l\cdot k_{l'} G_{ll'}-ik_l\cdot \varepsilon_{l'} \dot G_{ll'} +\varepsilon_l\cdot \varepsilon_{l'} \ddot G_{ll'}\Biggr)\Biggr\} 
\end{equation} 
whereas the vacuum expectation value of the auxiliary fields' part reads
\begin{eqnarray}
&&\Big\langle \prod_{l=1}^n {\cal V}_{c}(a_l;\tau_l)\Big \rangle = \Big\langle \prod_{l=1}^n {\cal V}_{c}(a_l;\tau_l)\Big \rangle_0\nonumber\\
&&+ {\rm tr}(T^{a_1}T^{a_2})  ~\left(2i\sin\frac{\theta}{2}\right)^{-2}~\Big\langle \prod_{l=3}^n {\cal V}_{c}(a_l;\tau_l)\Big \rangle_0+ \cdots \nonumber\\
&&+  {\rm tr}(T^{a_1}T^{a_2})  {\rm tr}(T^{a_3}T^{a_4}) \left({2i\sin\frac{\theta}{2}}\right)^{-4}~\Big\langle \prod_{l=5}^n {\cal V}_{c}(a_l;\tau_l)\Big \rangle_0 +\cdots \nonumber\\
&&+ \cdots 
\end{eqnarray}
where the different lines correspond to terms with  a different number of ``two-cycles" of auxiliary fields, and the subscript ``$0$" in a correlator indicates the absence of two-cycles therein. In each line the dots refers to the different ways of picking
the indices inside the traces. With ``two-cycles" we refer  to terms with two contractions among the same two pairs of fields, i.e.  $\langle \bar c_\alpha(\tau) c^\beta(\tau) \bar c_{\alpha'}(\tau') c^{\beta'}(\tau')\rangle$.~\footnote{Contractions among fields that belong to the same pair give no contribution as they set ${\rm tr}\, T^a=0$.}  Each of them comes with a term $\Delta(\tau-\sigma;\theta) \Delta(\sigma-\tau;\theta) = (2i\sin\frac{\theta}{2})^{-2}$.
In particular, for the two-gluon amplitude,  one gets
\begin{equation}
\Big\langle \prod_{l=1}^2 {\cal V}_{c}(a_l;\tau_l)\Big \rangle = {\rm tr} (T^{a_1}T^{a_2})\left(2i\sin\frac{\theta}{2}\right)^{-2}
\end{equation}
which is $\tau$-independent, and can be promptly integrated over $\theta$ to give the color factor
\begin{equation}
C^{ab}(r) = {\rm tr} (T^{a_1}T^{a_2})\int_0^{2\pi} \frac{d\theta}{2\pi} e^{is\theta} \left(2i\sin\frac{\theta}{2}\right)^{-2-N}
\label{eq:Cab}
\end{equation}
for the representation given by  a symmetric tensor product of $r$ fundamental representations of $SU(N)$, i.e. the representation described by the Young tableau 
\begin{equation}
\underbrace{\hskip0.1cm \yng(6) \hskip0.1cm }_r 
\label{eq:Young}
\end{equation}
By using the Wilson loop variable $z=e^{i\theta}$, expression~(\ref{eq:Cab}) can be computed as a contour integral 
\begin{equation}
C^{ab}(r) = {\rm tr} (T^{a_1}T^{a_2}) \oint_{\cal C} \frac{dz}{2\pi i}~\frac{z^{r+N}}{(z-1)^{N+2}}
\end{equation}
where the naive contour integral ${\cal C}$ is the unit circle centered in the origin $z=0$. However the latter integral may only have poles at the point $z=1$, that sits on the contour ${\cal C} $, and thus the contour integral must be defined through a suitable regularization. In fact, as shown in the~\ref{appA}, also the normalization factor for the path integral is obtained upon using a regularization, i.e. by shifting the angle of an infinitesimal negative imaginary part $\theta \to \theta-i\epsilon$. Applying the same regularization to the integral above corresponds to integrating the same function over a contour ${\cal C}_{\epsilon}$ that is a circle slightly larger than one, so that the integral precisely picks the pole at $z=1$. Hence,
\begin{equation}
C^{ab}(r) = {\rm tr} (T^{a}T^{b}) \binom{N+r}{N+1} = \delta^{a b} \frac12 \binom{N+r}{N+1}
\,,\quad n\geq 1
\end{equation} 
is the color factor for a scalar particle in a representation described by a symmetric tensor product of $r$ fundamental representations of $SU(N)$~\footnote{Notice that, $C(r)=\frac12 \binom{N+r}{N+1}$ is the correct normalization for generators $T^a_{(r)}$ in the representation~(\ref{eq:Young}), as it can also be derived directly from the well-known group theory identity $C(r) = C_2(r) d(r)/d(G)$, with $C_2(r)=\frac{r(N+r)(N-1)}{2N}$ being the quadratic Casimir, $d(G) =N^2-1$ the dimension of the group, and $d(r)=\binom{N+r-1}{r}$ the dimension of the representation.}.  Notice that for $r=0$, i.e. for a colorless particle, the contour has no simple pole, and thus vanishes, whereas for $r=1$ it yields $C^{ab}(1)=\frac12 \delta^{ab}$, which thus leads to the same result for the two-point function as found in~\cite{Bastianelli:2013pta}.
This  two-point function  gives the scalar contribution to the gluon self-energy. Let us explain it further.

By using integration by parts in the $\tau$ integrals, and defining $k:=k_1=-k_2$, the scalar contribution to the two-gluon amplitude
in four dimensions reads
\begin{align}
\Gamma_{\rm scal}(k_1,\varepsilon_1,a_1;k_2,\varepsilon_2,a_2)&=(2\pi)^4 \delta^{(4)}(k_1+k_2) C^{a_1a_2}(n) (\varepsilon_1\cdot \varepsilon_2 k^2-\varepsilon_1\cdot k \varepsilon_2\cdot k) ~\Pi(k^2)
\end{align}
where the dimensionally-extended form factor reads
\begin{align}
\Pi(k^2):=g^2\int_0^\infty\frac{dT}{T}\frac{e^{-Tm^2}}{(4\pi T)^{D/2}} T^2\int_0^1 d\tau_1 \int_0^1 d\tau_2 ~\dot G_{12}^2~e^{-Tk^2G_{12}}~.
\end{align}
The previous integral is UV logarithmically divergent (in $D=4$), and contributes to the renormalization of the gauge coupling constant, and to the associated $\beta$ function. It can be computed using dimensional regularization, i.e. $D=4-2\epsilon$  (and $\overline{\rm MS}$ scheme conventions) to give 
\begin{align}
\Pi(k^2) = \frac{g^2}{\epsilon 48\pi^2}-\frac{g^2}{16\pi^2} \int_0^1du ~(1-2u)^2 \log\frac{M^2(u,k^2)}{\mu^2} +{\cal O}(\epsilon)
\end{align}   
where $M^2(u,k^2) = m^2 +k^2 u(1-u)$. In the UV limit $k^2 >> m^2$ the finite part of the previous expression reduces to
\begin{align}
\Pi_{UV}(k^2) = -\frac{g^2}{48\pi^2} \log\frac{k^2}{\mu^2}~.
\end{align}
Furthermore, the tensor structure can be written in terms of the (abelian part of the) QCD lagrangian, if the gauge field is represented as the sum of two gluons $A^\mu(x) =\varepsilon^\mu_1 T^{a_1}e^{ik_1\cdot x} +\varepsilon^\mu_2 T^{a_2}e^{ik_2\cdot x} $. By rescaling the coupling constant in the gauge field one gets (a similar rescaling takes place in $\Pi(k^2)$ above)
\begin{align}
-\frac{1}{2g^2}\int d^4 x ~{\rm tr}(F_{\mu\nu} F^{\mu\nu} )^{(\rm abe)}(x) =-\frac{2}{g^2}(2\pi)^4 \delta^{(4)}(k_1+k_2) (k^2 \varepsilon_1\cdot \varepsilon_2-\varepsilon_1\cdot k \varepsilon_2\cdot k) C^{a_1a_2}(1)
\end{align} 
or
\begin{align}
-\frac{1}{2g^2}{\rm tr}(F_{\mu\nu} F^{\mu\nu} )^{(\rm abe)}(k) = -\frac{2}{g^2} (k^2 \varepsilon_1\cdot \varepsilon_2-\varepsilon_1\cdot k \varepsilon_2\cdot k) C^{a_1a_2}(1)~.
\end{align}
Hence,  the renormalized two-point function can be obtained from 
\begin{align}
\tilde \Gamma^{(2)}(k,-k) = -\Biggl[\frac{1}{g^2}+\binom{N+r}{N+1}\frac{1}{96\pi^2} \log\frac{k^2}{\mu^2}\Biggr] ~\frac{1}{2} {\rm tr}(F_{\mu\nu} F^{\mu\nu} )(k)
\end{align}
where a term $(2\pi)^4 \delta^{(4)}(k_1+k_2)$ was stripped off. The running coupling constant can be read off from the latter
\begin{align}
g^2(\mu) = g^2\Biggl[1+g^2\binom{N+r}{N+1}\frac{1}{96\pi^2} \log\frac{k^2}{\mu^2}\Biggr]^{-1}
\end{align}
from which the $\beta$ function can be computed
\begin{align}
\beta(g) =\frac{\partial g(\mu)}{\partial \log \mu} =\binom{N+r}{N+1}\frac{g^3}{96\pi^2} = \frac13 C(r)\frac{g^3}{16\pi^2}  ~.
\end{align}

\section{The spin 1/2 particle with color degrees of freedom}
The generalization of the results described above to a spin $1/2$ colored fermion, involves the (local) ${\mathcal N}=1$ supersymmetrization of the worldline action~\eqref{eq:W-action}. This is achieved by adding fermionic coordinates $\psi^\mu$, and a gravitino $\chi$. The massless version of the action reads
\begin{align}
S[x,\psi,c,\bar c,e,\chi,a;A]= \int_0^1d\tau & \Biggl[ \frac{1}{2e}(\dot x^\mu-\chi\psi^\mu)^2 +\frac{1}{2}\psi_\mu\dot \psi^\mu +\bar c^\alpha (\partial_\tau+ia)c_\alpha -isa
\nonumber\\&-ig \dot x^\mu A^a_\mu\, \bar c^\alpha (T^a)_\alpha{}^\beta c_\beta+\frac{ie}{2}g \psi^\mu \psi^\nu F^a_{\mu\nu} \bar c^\alpha (T_a)_\alpha{}^\beta c_\beta\Biggr]
\end{align} 
which, as above, can be used on a circle path integral to compute the one-loop effective action. In order to do so one has to gauge-fix the worldline gauge fields, and divide out the volume of the gauge group. The gravitino $\chi$ is taken with antiperiodic boundary conditions and can be gauged-away completely.  Finally one can twist the auxiliary field to absorb the constant $U(1)$ field, and write the nonabelian gauge field as a sum of gluons (a mass term can be added as well). This allows to read off the gluon vertex operators, that in turn allow to compute gluon amplitudes as worldline path integral averages of such vertex operators. The computation of the two-gluon amplitudes proceeds in the very same way as for the scalar case. One gets
\begin{align}
\Gamma_{\rm spin}(k_1,\varepsilon_1,a_1;k_2,\varepsilon_2,a_2)&=(2\pi)^4 \delta^{(4)}(k_1+k_2) C^{a_1a_2}(r) (\varepsilon_1\cdot \varepsilon_2 k^2-\varepsilon_1\cdot k \varepsilon_2\cdot k) ~\Pi(k^2)
\end{align} 
where now
\begin{align}
\Pi(k^2) =-2^{D/2}g^2\int_0^\infty \frac{dT}{2T}\frac{e^{-Tm^2}}{(4\pi T)^{D/2}}T^2\int_0^1d\tau_1 \int_0^1d\tau_2~e^{-Tk^2 G_{12}}\Big( \dot G_{12}^2-4S_{12}^2 \Big)
\end{align}
with $S(\tau- \sigma)=\frac12 \epsilon(\tau -\sigma)$ being the fermionic propagator. Using dimensional regularization the previous integral reduces to (up to irrelevant constants)
\begin{align}
\Pi(k^2) =\frac{g^2}{\epsilon 12\pi^2}-\frac{g^2}{2\pi^2} \int_0^1du~ u(1-u) \log\frac{M^2(u,k^2)}{\mu^2} +{\cal O}(\epsilon)
\end{align}
which in the UV limit gives $\Pi_{UV}(k^2) = -\frac{g^2}{12\pi^2} \log\frac{k^2}{\mu^2} $, and the fermionic contribution to the beta function thus reads
\begin{align}
\beta(g) =\frac{\partial g(\mu)}{\partial \log \mu} =\binom{N+r}{N+1}\frac{g^3}{24\pi^2} = \frac43 C(r)\frac{g^3}{16\pi^2}  
\end{align}
that reproduces well known results; see e.g.~\cite{Peskin:1995ev}.

\section{Conclusions and outlook}
A colored relativistic particle action is used to compute the scalar contribution to the one-loop gluon effective action, and commuting auxiliary fields are used to implement the path ordering and to reproduce the desired representation for the scalar particle. The main result is a Bern-Kosower like master formula for scalar contribution to the one-loop $n$-gluon amplitude. The present manuscript complements an earlier publication where the case of anticommuting auxiliary fields was described. 

Within this framework a natural extension concerns the worldline representation of the scalar propagator ``dressed" with external gluons. Such dressed propagator describes the sum of all tree-level diagrams with one incoming and one outgoing scalar particle, and an arbitrary number of gluons, and it is related to the above particle path integral defined on a line, i.e. taken with fixed (Dirichlet) initial and final boundary conditions, both for the coordinates and for the auxiliary variables. This last condition is necessary in order to assign color states to the scalar particles. Although such computation should be relatively straightforward, the description of  the colored particle with Dirichlet boundary conditions surely  deserves a  deeper investigation, and we leave it for future studies. 

\section*{Acknowledgments}{The work of OC was partly supported by the UCMEXUS-CONACYT grant CN-12-564. He is grateful to Idrish Huet and Christian Schubert for helpful discussions. }EL acknowledges partial support from SNF Grant No. 200020-149150/1.

\appendix
\section{Path integral normalization and propagator for the auxiliary fields}
\label{appA}
The path integral normalization for a pair of twisted auxiliary fields satisfying the conditions $c(1)=e^{i\theta} c(0)$ and $\bar c(1)=e^{-i\theta} \bar c(0)$, corresponds to the normalization for a coherent state path integral of a harmonic oscillator of frequency $\theta$. It reads
\begin{eqnarray}
&&\Big[{\rm Det}_{TBC} (\partial_\tau)\Big]^{-1} =\Big[{\rm Det}_{PBC} (\partial_\tau +i \theta)\Big]^{-1}=\int_{PBC} D\bar c D c~e^{-\int_0^1 \bar c (\partial_\tau +i \theta )c} \nonumber\\
 &&= {\rm Tr}~e^{-i\frac{\theta}{2}(\hat c^\dagger \hat c +\hat c \hat c^\dagger)}= \sum_{n=0}^\infty e^{-i\theta(n+\frac 12)} = \frac{1}{2i\sin\frac{\theta}{2}}
\end{eqnarray}
where a fast-oscillating exponential was regulated away by shifting the angle of an infinitesimal, negative, imaginary part $\theta \to \theta-i\epsilon$. 
Alternatively one may perform a direct calculation of the determinant, getting
\begin{eqnarray}
&& \Big[{\rm Det}_{TBC} (\partial_\tau)\Big]^{-1} =\left[\frac{{\rm Det}_{PBC} (\partial_\tau +i \theta))}{{\rm Det}'_{PBC} (\partial_\tau))} \right]^{-1}\frac{1}{{\rm Det}'_{PBC} (\partial_\tau))}\nonumber\\
&& =\left[ \frac{\prod_n i(2\pi n +\theta)}{\prod_n' i2\pi n}\right]^{-1}\cdot 1 = \left[i\theta \prod_{n=1}^\infty \left(1-\left(\frac{\theta}{2\pi n}\right)^2 \right)\right]^{-1} =\frac{1}{2i\sin\frac{\theta}{2}}
\end{eqnarray}
where, in the last equality we used the infinite product expansion for $\frac{\sin x}{x}$. On the other hand for the propagator we have that
\begin{equation}
\langle c^\alpha(\tau) \bar c_\beta (\sigma)\rangle = \delta^\alpha_\beta \Delta(\tau-\sigma;\theta)
\end{equation}
with
\begin{eqnarray*}
&&\partial_\tau \Delta(\tau-\sigma;\theta) = \delta(\tau-\sigma)
\nonumber\\
&&\Delta(1-\sigma;\theta) = e^{i\theta} \Delta(0-\sigma;\theta)~,
\end{eqnarray*}
and
\begin{equation}
\Delta(\tau-\sigma;\theta)=\frac{1}{2i\sin\frac{\theta}{2}} \Biggl[ e^{i\frac\theta 2}\theta(\tau-\sigma) + e^{-i\frac\theta 2}\theta(\sigma-\tau)\Biggr]~.
\end{equation}
Notice that, by taking $\theta=\phi+\pi$, the latter is mapped onto the propagator of~\cite{Bastianelli:2013pta}, for auxiliary fields with boundary conditions $c(1)=-e^{i\theta} c(0)$ and $\bar c(1)=-e^{-i\theta} \bar c(0)$.

\section*{References}

\end{document}